\documentclass[%
 aip,
 jmp,%
 amsmath,amssymb,
 reprint,%
]{revtex4-1}

\usepackage{graphicx}
\usepackage{dcolumn}
\usepackage{bm}
\usepackage[cal=boondox]{mathalfa}

\DeclareMathAlphabet\mathbfcal{OMS}{cmsy}{b}{n}

\begin{document}

\preprint{AIP/123-QED}

\title[Relativistic Nonlinear Whistler Waves ...]{Relativistic Nonlinear Whistler Waves in Cold Magnetized Plasmas}


\author{Stephan I. Tzenov} 

\email{stephan.tzenov@eli-np.ro} 

\affiliation{Extreme Light Infrastructure - Nuclear Physics, 077125 Magurele, Bucharest Ilfov County, Romania}



\date{\today}

\begin{abstract}
Starting from the Vlasov-Maxwell equations describing the dynamics of various species in a quasi-neutral plasma immersed in an external solenoidal magnetic field and utilizing a technique known as the hydrodynamic substitution, a relativistic hydrodynamic system of equations governing the dynamics of various species has been obtained.

Based on the method of multiple scales, a system comprising three nonlinear Schrodinger equation for the transverse envelopes of the three basic whistler modes, has been derived. 

Using the method of formal series of Dubois-Violette, a traveling wave solution of the derived set of coupled nonlinear Schrodinger equations in both the relativistic and the non relativistic case has been obtained. 

An intriguing feature of our description is that whistler waves do not perturb the initial uniform density of plasma electrons. The plasma response to the induced whistler waves consists in transverse velocity redistribution, which follows exactly the behaviour of the electromagnetic whistlers. This property may have an important application for transverse focusing of charged particle beams in future laser plasma accelerators. Yet another interesting peculiarity are the selection rules governing the nonlinear mode coupling. According to these rules self coupling between modes in the non relativistic regime is absent, which is a direct consequence of the vector character of the interaction governed by the Lorentz force.

\end{abstract}

\maketitle

\section{\label{sec:intro}Introduction} 

Whistler waves are probably one of the first plasma waves observed and studied for more than a century. They were apparently discovered in 1894 as noises resembling a whistle, heard by telephone operators. It seems that these noises were induced by the electromagnetic fields of whistler waves. These waves propagate parallel to the applied magnetic field being circularly polarized in a plane transverse to the direction of propagation. Since their discovery in the end of 19-th century, it has become conventional in the physics of magnetized plasmas to call such structures waves in the whistler mode.

The first analytical approach to the linear dispersion properties of whistler waves is possibly the one suggested by Appleton \cite{Appleton} and Hartree \cite{Hartree}, who proposed the famous Appleton-Hartree dispersion equation. This equation was originally derived from Lorentz classical equations of motion of electrons in an electromagnetic field. A vast body of literature exists on whistlers and related phenomena in at least three areas: space plasmas, laboratory plasmas, and solid state physics. Since it is impossible to refer to all this important work, we shall restrict ourselves in mentioning some of the early and relatively recent approaches to the linear and nonlinear whistler waves. 

While the linear stability properties of the electromagnetic waves in the whistler mode are relatively well studied \cite{Weibel,Neufeld,Bell,Sudan}, there is a serious gap in the understanding of their nonlinear behaviour. The general theory of nonlinear waves in a cold, collisionless relativistic plasma with stationary ions was initiated by Akhiezer and Polovin \cite{Akhiezer}. They established a correspondence between the wave motion of the plasma and the motion of a non relativistic particle in a certain potential field. Among the older papers on the subject the one by Chen et al. \cite{Chen} should be mentioned. These authors have obtained a class of exact large-amplitude traveling-wave solutions to the nonlinear Vlasov-Maxwell equations describing a one-dimensional collisionless magnetized plasma. In addition, it has been shown that these waves are complementary to the electrostatic Bernstein-Greene-Kruskal modes \cite{BGK} and can be classified as nonlinear fast electromagnetic waves on one hand, and (slow) electromagnetic whistler waves on the other. While the BGK modes are longitudinal, the whistler modes are transverse, in other words, the components of the electric and magnetic field of the whistler wave parallel to the external magnetic field are both zero. It seems that the study of the nonlinear behaviour of whistler waves has been initiated by Taniuti and Washimi \cite{Taniuti}, who obtained a nonlinear Schrodinger equation for the slowly varying wave amplitude. 

Recently, a fully nonlinear theory for stationary whistler waves propagating parallel to the ambient magnetic field in a cold plasma has been developed by Dubinin et al \cite{Dubinin}. The resonant acceleration of electrons via Landau and fundamental
cyclotron resonances by a oblique whistler wave using a single particle Hamiltonian technique has been studied by Artemyev et al \cite{Artemyev}. The dispersion of linear waves in a uniform cold quantum plasma immersed in an external axial magnetic field has been derived recently \cite{Ren}. The dispersion relation thus obtained, can be viewed as a quantum generalization of the classical Appleton-Hartree equation. It has been also shown that the dispersion of the Langmuir wave becomes whistler-like due to quantum effects.

The present article is devoted to the analysis of nonlinear waves and coherent structures in the whistler mode, which build up and propagate in classical cold plasmas. Coherent structures which result from the nonlinear interaction between plane waves evolve on time and/or spatial scales comparatively large compared to those the fast oscillations occur. 

The article is organized as follows. In Section \ref{sec:basic}, we review the physical principles and the underlying equations on which the subsequent exposition is based on, including the derivation of the cold hydrodynamic picture by using the so-called hydrodynamic substitution. Using the method of multiple scales, we perform in Section \ref{sec:multiple} a reduction of the macroscopic fluid equations coupled to the wave equations for the self-consistent electromagnetic fields. As a result, we obtain a system of coupled nonlinear Schrodinger equation describing the evolution of the slowly varying amplitudes of the three basic whistler wave modes. An interesting feature of this system is that in the non relativistic regime self coupling between the basic modes is not allowed, which is due to the special form of the matrix of coupling coefficients. This represents a sort of a selection rule, according to which a generic mode cannot couple with itself -- a feature, which is a consequence of the vector character of the nonlinear coupling between modes. An approximate traveling wave solution of the coupled nonlinear Schrodinger equations in both the relativistic and the non relativistic case has been found in Section \ref{sec:solcnse}. Finally, in Section \ref{sec:conclude}, we draw some conclusions. 

\section{\label{sec:basic}Theoretical Model and Basic Equations}

We start with the description of a plasma comprised of electrons and ions in an external constant magnetic field ${\bf B}_0 = B_0 {\bf e}_x$, where ${\bf e}_x = {\left( 1, 0, 0 \right)}$ is the unit vector in the $x$-direction. The Hamiltonian describing the dynamics of the different species (electrons and ions) labelled by the subscript $a$ can be written as
\begin{equation}
{\cal H}_a = c {\sqrt{m_a^2 c^2 + {\left( {\widetilde{\bf p}} - q_a {\bf A}_0 - q_a {\bf A}  \right)}^2}} + q_a \Phi, \label{HamilA}
\end{equation}
where $m_a$ and $q_a$ are the rest mass and the charge of a particle of species $a$, respectively, $c$ is the speed of light in vacuum, ${\widetilde{\bf p}}$ is the particle's canonical momentum, ${\bf A}_0$ is the electromagnetic vector potential responsible for the external axial magnetic field, and ${\bf A}$ and $\Phi$ are the electromagnetic potentials of the self-fields produced by the plasma particles. 

Introducing the kinetic momenta 
\begin{equation}
{\bf p} = {\widetilde{\bf p}} - q_a {\bf A}_0 - q_a {\bf A}, \label{KineticMom}
\end{equation}
we can write the Hamilton's equations of motion as 
\begin{equation}
{\frac {{\rm d}{\bf x}} {{\rm d} t}} = {\frac {\bf p} {m_a \gamma_a}}, \qquad \quad {\frac {{\rm d}{\bf p}} {{\rm d} t}} = q_a {\left[ {\bf E} + {\frac {{\bf p}} {m_a \gamma_a}} \times {\left( {\bf B}_0 + {\bf B} \right)} \right]}. \label{HamilEqMot}
\end{equation}
Here 
\begin{equation}
\gamma_a {\left( {\bf p} \right)} = {\frac {1} {m_a c}} {\sqrt{m_a^2 c^2 + {\bf p}^2}}, \label{GammaA}
\end{equation}
is the Lorentz gamma factor, while the self-consistent electric ${\bf E}$ and magnetic ${\bf B}$ fields are expressed in the usual way 
\begin{equation}
{\bf E} = - {\boldsymbol{\nabla}} \Phi - \partial_t {\bf A}, \qquad \qquad {\bf B} = {\boldsymbol{\nabla}} \times {\bf A}, \label{SelfFields}
\end{equation}
where ${\boldsymbol{\nabla}} = {\left( \partial_x, \partial_y, \partial_z \right)}$ is the well-known gradient operator. In addition, $\partial_t$ implies partial derivative with respect to the time, while $\partial_{x,y,z}$ denotes partial differentiation with respect to the spatial variables.

The Vlasov equation for the one-particle distribution function $f_a {\left( {\bf x}, {\bf p}; t \right)}$ describing the statistical evolution of species of type $a$ reads as 
\begin{equation}
\partial_t f_a + {\bf v} \cdot {\boldsymbol{\nabla}} f_a + q_a {\left[ {\bf E} + {\bf v} \times {\left( {\bf B}_0 + {\bf B} \right)} \right]} \cdot {\boldsymbol{\nabla}}_p f_a = 0, \label{VlasovA}
\end{equation}
where 
\begin{equation}
{\bf v} = {\frac {\bf p} {m_a \gamma_a}}, \label{Velocity}
\end{equation}
and ${\boldsymbol{\nabla}}_p = {\left( \partial_{p_x}, \partial_{p_y}, \partial_{p_z} \right)}$. The self-consistent electromagnetic potentials ${\bf A}$ and $\Phi$ satisfy the wave equations 
\begin{equation}
{\boldsymbol{\Box}} {\bf A} = - \mu_0 \sum \limits_a q_a \int {\frac {{\rm d}^3 {\bf p} {\bf p}} {m_a \gamma_a}} f_a {\left( {\bf x}, {\bf p}; t \right)}, \label{WaveEqA}
\end{equation}
\begin{equation}
{\boldsymbol{\Box}} \Phi = - {\frac {1} {\epsilon_0}} \sum \limits_a q_a \int {\rm d}^3 {\bf p} f_a {\left( {\bf x}, {\bf p}; t \right)}, \label{WaveEqPhi}
\end{equation}
where $\epsilon_0$ and $\mu_0$ are the electric permittivity and the magnetic permeability of free space, respectively, and
\begin{equation}
{\boldsymbol{\Box}} = {\boldsymbol{\nabla}}^2 - {\frac {1} {c^2}} \partial_t^2, \label{dAlembert}
\end{equation}
is the d'Alembert operator. Instead of using Eq. (\ref{WaveEqPhi}) for the calculation of the electric field according to the first of Eqs. (\ref{SelfFields}), the Lorentz gauge 
\begin{equation}
{\frac {1} {c^2}} \partial_t \Phi + {\boldsymbol{\nabla}} \cdot {\bf A} = 0, \label{LorentzGauge}
\end{equation}
will be utilized in the sequel. More precisely, the relation  
\begin{equation}
\partial_t {\bf E} = c^2 {\left[ {\boldsymbol{\nabla}} {\left( {\boldsymbol{\nabla}} \cdot {\bf A} \right)} - {\frac {1} {c^2}} \partial_t^2 {\bf A} \right]}, \label{EARelation}
\end{equation}
will be used. 

It will prove convenient and useful for the subsequent exposition to introduce dimensionless variables according to the relations  
\begin{equation}
{\widetilde{t}} = \omega_{pe} t, \qquad {\widetilde{\bf x}} = {\frac {\omega_{pe}} {c}} {\bf x}, \qquad {\widetilde{\bf p}} = {\frac {\bf p} {m_e c}}, \label{Dimensionless1}
\end{equation}
\begin{equation}
{\widetilde{\bf A}} = {\frac {e {\bf A}} {m_e c}}, \qquad \varphi = {\frac {e \Phi} {m_e c^2}}, \qquad {\widetilde{\bf v}} = {\frac {\bf v} {c}} \label{Dimensionless2}
\end{equation}
\begin{equation}
{\widetilde{\bf E}} = {\frac {e {\bf E}} {m_e c \omega_{pe}}}, \qquad {\widetilde{\bf B}} = {\frac {e {\bf B}} {m_e \omega_{pe}}}, \qquad {\widetilde{f}}_a = {\frac {m_e^3 c^3} {n_a}} f_a, \label{Dimensionless3}
\end{equation}
where $n_a$ is the volumetric number density of the species of type $a$, and 
\begin{equation}
\omega_{pe}^2 = {\frac {e^2 n_e} {\epsilon_0 m_e}}, \label{ElPlasFreq}
\end{equation}
is the electron plasma frequency. To simplify notations, the tilde-signs above the new dimensionless variables will be omitted in the sequel. Thus, the dimensionless Vlasov-Maxwell system of equations can be written in the form 
\begin{equation}
\partial_t f_a + {\bf v} \cdot {\boldsymbol{\nabla}} f_a - \nu_a {\bf e}_x \times {\bf v} \cdot {\boldsymbol{\nabla}}_p f_a + Z_a {\left( {\bf E} + {\bf v} \times {\bf B} \right)} \cdot {\boldsymbol{\nabla}}_p f_a = 0, \label{VlasovDim}
\end{equation}
\begin{equation}
{\boldsymbol{\Box}} {\bf A} = - \sum \limits_a \lambda_a \int {\rm d}^3 {\bf p} {\bf v} f_a {\left( {\bf x}, {\bf p}; t \right)}, \label{WaveEqADim}
\end{equation}
\begin{equation}
{\boldsymbol{\Box}} \varphi = - \sum \limits_a \lambda_a \int {\rm d}^3 {\bf p} f_a {\left( {\bf x}, {\bf p}; t \right)}, \label{WaveEqPhiDim}
\end{equation}
\begin{equation}
\partial_t {\bf E} = {\boldsymbol{\nabla}} {\left( {\boldsymbol{\nabla}} \cdot {\bf A} \right)} - \partial_t^2 {\bf A}, \qquad \qquad {\bf B} = {\boldsymbol{\nabla}} \times {\bf A}, \label{EARelatDim}
\end{equation}
where the d'Alembert operator in the new dimensionless variables reads as  
\begin{equation}
{\boldsymbol{\Box}} = {\boldsymbol{\nabla}}^2 - \partial_t^2. \label{dAlembertDim}
\end{equation}
The quantity $\lambda_a$ in Eqs. (\ref{WaveEqADim}) and (\ref{WaveEqPhiDim}) is defined according to the expressions 
\begin{equation}
\lambda_a = {\frac {Z_a n_a} {n_e}}, \label{QuantMuLa}
\end{equation}
where $Z_a$ is the charge state ${\left( q_a = Z_a e \right)}$ of the particles of the species $a$. In addition, 
\begin{equation}
\nu_a = {\frac {\omega_a} {\mu_a \omega_{pe}}}, \qquad \qquad \omega_a = {\frac {q_a B_0} {m_a}}, \label{CycFreq}
\end{equation}
where $\omega_a$ is the cyclotron frequency of particles of type $a$, and $\nu_a$ is the corresponding scaled cyclotron frequency with respect to the electron plasma frequency. Moreover, 
\begin{equation}
\mu_a = {\frac {m_e} {m_a}}, \label{RelatMass}
\end{equation}
denotes the inverse relative rest mass of the plasma species of type $a$ with respect to the electron rest mass, which for protons and ions is a small quantity. 

Consider the so-called hydrodynamic substitution \cite{Silin} 
\begin{equation}
f_a {\left( {\bf x}, {\bf p}; t \right)} = \varrho_a {\left( {\bf x}; t \right)} {\boldsymbol{\delta}}^3 {\left[ {\bf p} - {\frac {1} {\mu_a}} \gamma_a {\left( {\bf x}; t \right)} {\bf v}_a {\left( {\bf x}; t \right)} \right]}. \label{HydDynSubst}
\end{equation}
The physical meaning of this substitution lies in the fact that a unique velocity ${\bf v}_a {\left( {\bf x}; t \right)}$ corresponds to each point ${\bf x}$ in configuration space, where the hydrodynamic density distribution $\varrho_a {\left( {\bf x}; t \right)}$ is defined. Substituting the above expression (\ref{HydDynSubst}) into the Vlasov equation (\ref{VlasovDim}) and taking into account Eqs. (\ref{WaveEqADim}) and (\ref{WaveEqPhiDim}), we obtain the cold hydrodynamic equations 
\begin{equation}
\partial_t \varrho_a + {\boldsymbol{\nabla}} \cdot {\left( \varrho_a {\bf v}_a \right)} = 0, \label{Continuitya}
\end{equation}
\begin{equation}
\partial_t {\left( \gamma_a {\bf v}_a \right)} + {\bf v}_a \cdot {\boldsymbol{\nabla}} {\left( \gamma_a {\bf v}_a \right)} + {\bar{\omega}}_a {\bf e}_x \times {\bf v}_a  = \mu_a Z_a {\left( {\bf E} + {\bf v}_a \times {\bf B} \right)}, \label{MomBalancea}
\end{equation}
supplemented with the equations for the self-fields 
\begin{equation}
{\boldsymbol{\Box}} {\bf A} = - \sum \limits_a \lambda_a \varrho_a {\bf v}_a, \qquad \quad {\boldsymbol{\Box}} \varphi = - \sum \limits_a \lambda_a \varrho_a, \label{WaveEqsHyd}
\end{equation}
where the notation ${\bar{\omega}}_a = \omega_a / \omega_{ep}$ has been introduced. 

The system of hydrodynamic equations and wave equations for the self-fields (\ref{Continuitya}) -- (\ref{WaveEqsHyd}) together with Eq. (\ref{EARelatDim}) will be the starting point of our subsequent analysis. 

\section{\label{sec:multiple}Multiple Scales Reduction of the Hydrodynamic and Field Equations}

The frequency of the electromagnetic wave excited in our plasma will be considered much higher than the (relative) ion-cyclotron frequency ${\bar{\omega}}_i$. Since $\mu_a \ll 1$ for $a \neq e$ (formally, $\mu_e = 1$), we can further neglect the ion motion and take into account the contribution coming from the much lighter electrons only. The only role of the immobile heavy plasma species is to provide a charge neutralizing environment, which will be used implicitly throughout the rest of the exposition. In addition, we shall assume that the plasma wave properties are anisotropic, implying that the longitudinal and the transverse plasma waves depend on the longitudinal (in the direction of the applied external magnetic field ${\bf B}_0$) $x$-coordinate only.

For the sake of simplicity of notation the bar over ${\bar{\omega}}_e$ as well as the subscript $e$ on all hydrodynamic variables referring to the plasma electrons will be omitted. In addition, the simplified notations and the new complex valued variables 
\begin{equation}
v = v_x, \qquad V = v_y + i v_z, \qquad E = E_x, \label{NewVariab1}
\end{equation}
\begin{equation}
{\cal E} = E_y + i E_z, \qquad A = A_x, \qquad {\cal A} = A_y + i A_z, \label{NewVariab2}
\end{equation}
\begin{equation}
B_x = 0, \qquad \qquad {\cal B} = B_y + i B_z = i \partial_x {\cal A}, \label{NewVariab3}
\end{equation}
will prove convenient in the subsequent exposition. Thus, the basic equations to be solved perturbatively can be written as 
\begin{equation}
\partial_t \varrho + \partial_x {\left( \varrho v \right)} = 0, \label{Continuity}
\end{equation}
\begin{equation}
\partial_t {\left( \gamma v \right)} + v \partial_x {\left( \gamma v \right)} = - E - {\frac {1} {2}} {\left( V \partial_x {\cal A}^{\ast} + V^{\ast} \partial_x {\cal A} \right)}, \label{MomBalancev}
\end{equation}
\begin{equation}
\partial_t {\left( \gamma V \right)} + i \omega_e V + v \partial_x {\left( \gamma V \right)} = \partial_t {\cal A} + v \partial_x {\cal A}, \label{MomBalanceV}
\end{equation}
\begin{equation}
{\boldsymbol{\Box}} A = \varrho v, \qquad \quad {\boldsymbol{\Box}} {\cal A} = \varrho V, \qquad \quad \partial_t E = {\boldsymbol{\Box}} A. \label{WaveEqsGen}
\end{equation}

Following the standard procedure of the multiple scales reduction method \cite{TzenovBOOK,Nayfeh,Kevorkian} applied to the system of equations (\ref{Continuity}) -- (\ref{WaveEqsGen}), we represent the electron density distribution $\varrho$, the electron current velocities $v$ and $V$, the longitudinal electric field $E$ and the components $A$ and ${\cal A}$ of the electromagnetic vector potential as a perturbation expansion according to the expressions 
\begin{equation} 
\varrho = 1 + \sum \limits_{k=1}^{\infty} \epsilon^k \varrho_k, \qquad v = \sum \limits_{k=1}^{\infty} \epsilon^k v_k, \qquad V = \sum \limits_{k=1}^{\infty} \epsilon^k V_k, \label{Expand1}
\end{equation}
\begin{equation} 
E = \sum \limits_{k=1}^{\infty} \epsilon^k E_k, \qquad A = \sum \limits_{k=1}^{\infty} \epsilon^k A_k, \qquad {\cal A} = \sum \limits_{k=1}^{\infty} \epsilon^k {\cal A}_k. \label{Expand2}
\end{equation}
Here the quasi-neutrality condition sustained by the immobile heavy plasma background has been made use of implicitly, since the only nonzero quantity in the stationary solution of Eqs. (\ref{Continuitya}) -- (\ref{WaveEqsHyd}) is $\varrho_0 = 1$. In addition, $\epsilon$ is a formal small parameter, which will be set equal to one at the end of all calculations. The differential operators with respect to the time $t$ and to the longitudinal spatial variable $x$ are also expanded in the small parameter $\epsilon$ as follows
\begin{equation}
\partial_t = \sum \limits_{n=0}^{\infty} \epsilon^n \partial_{t_n}, \qquad \qquad \partial_x = \sum \limits_{n=0}^{\infty} \epsilon^n \partial_{x_n}, \label{PerturbOper}
\end{equation}
where 
\begin{equation}
t_n = \epsilon^n t, \qquad \qquad x_n = \epsilon^n x. \label{PerturbVar}
\end{equation}
The next step consists in expanding the system of hydrodynamic and field equations (\ref{Continuity}) -- (\ref{WaveEqsGen}) in the formal small parameter $\epsilon$. Their perturbation solution will be obtained order by order together with a reduction procedure aimed at elimination of secular terms (starting with second order). As is usually the case, the third order of such reduction procedure yields the sought for nonlinear amplitude equations for the slowly varying envelopes. 

The general form of the $n$-th order perturbation equations to be used in what follows can be written as 
\begin{equation}
\partial_t \varrho_n + \partial_x v_n + R_n = 0, \label{PertContn}
\end{equation}
\begin{equation}
\partial_t v_n = - E_n - W_n, \qquad {\widehat{\mathbfcal L}} V_n = \partial_t {\cal A}_n + U_n, \label{PertMomBalvVn}
\end{equation}
\begin{equation}
{\boldsymbol{\Box}} A_n = v_n + N_n, \qquad \quad {\boldsymbol{\Box}} {\cal A}_n = V_n + S_n,  \label{PertWEqsn}
\end{equation}
\begin{equation}
\partial_t E_n = {\boldsymbol{\Box}} A_n + {\cal D}_n, \label{PertWEqEn}
\end{equation}
where the operator ${\widehat{\mathbfcal L}}$ is given by the expression 
\begin{equation}
{\widehat{\mathbfcal L}} = \partial_t + i \omega_e. \label{LOperator}
\end{equation}
Here, the auxiliary perturbation quantities $R_n$, $W_n$, $U_n$, $N_n$, $S_n$ and ${\cal D}_n$ involve the hydrodynamic and electromagnetic field variables already calculated in previous orders. Manipulating the above Eqs. (\ref{PertMomBalvVn}) -- (\ref{PertWEqEn}) in an obvious manner, we arrive at the basic equations  
\begin{equation}
{\boldsymbol{\Box}} {\left( \partial_t^2 + 1 \right)} A_n = \partial_t^2 N_n - \partial_t W_n - {\cal D}_n, \label{BasicLongn}
\end{equation}
\begin{equation}
{\left( {\boldsymbol{\Box}} {\widehat{\mathbfcal L}} - \partial_t \right)} {\cal A}_n = {\widehat{\mathbfcal L}} S_n + U_n, \label{BasicWhistlern}
\end{equation}
to be solved order by order. 

\subsection{\label{subsec:first}First Order}

Since (by definition) all contributions from $R_1$, $W_1$, $U_1$, $N_1$, $S_1$ and ${\cal D}_1$ are equal to zero, we must solve the homogeneous Eqs. (\ref{BasicLongn}) for the longitudinal plasma waves and (\ref{BasicWhistlern}) for the transverse ones. The latter describe the linear wave properties of our system. 

\subsubsection{Longitudinal Waves}

Equation (\ref{BasicLongn}) for the longitudinal component of the electromagnetic vector potential has a trivial solution 
\begin{equation}
A_1 = 0, \label{ALong1}
\end{equation}
in first order. This automatically implies that all the rest quantities of interest related to the longitudinal degree of freedom also vanish 
\begin{equation}
E_1 = 0, \qquad v_1 = 0, \qquad \varrho_1 = 0. \label{EVRhoLong1}
\end{equation}

\subsubsection{Whistler Waves}

The standard procedure of seeking wave solutions of the form $\sim {\rm e}^{i {\left( k x - \omega t \right)}}$ to the equation 
\begin{equation}
{\left( {\boldsymbol{\Box}} {\widehat{\mathbfcal L}} - \partial_t \right)} {\cal A}_1 = 0, \label{BasicWhistler1}
\end{equation}
yields a dispersion equation 
\begin{equation}
D {\left( k, \omega \right)} = \omega - \Box_{\omega} {\cal L}_{\omega} = 0, \label{DisperEq}
\end{equation}
for the transverse whistler waves. Here 
\begin{equation}
\Box_{\omega} = \omega^2 - k^2, \qquad \qquad {\cal L}_{\omega} = \omega - \omega_e. \label{DispConst}
\end{equation}
It can be easily verified that for typical values of the electron-cyclotron frequency $\omega_e$ the dispersion equation (\ref{DisperEq}) possesses three distinct real roots $\omega_n {\left( k \right)}$, where $n = 1, 2, 3$. Thus, the general solution of Eq. (\ref{BasicWhistler1}) can be written as 
\begin{equation}
{\cal A}_1 = \sum \limits_{n=1}^3 {\cal C}_n {\rm e}^{i \psi_n}, \qquad \quad \psi_n = k x - \omega_n t. \label{SolAW1}
\end{equation}
An important note is now in order. The arbitrary to this end wave amplitudes ${\cal C}_n$ are constants with respect to the fast time $t$ and length $x$ scales, but they can depend on the slower ones $t_1, t_2, \dots$ and $x_1, x_2, \dots$ in general. Separating real from imaginary part in Eq. (\ref{SolAW1}), we can express the transverse components $A_{y1}$ and $A_{z1}$ of the vector potential as follows 
\begin{equation}
A_{y1} = {\frac {1} {2}} \sum \limits_{n=1}^3 {\left( {\cal C}_n {\rm e}^{i \psi_n} + {\cal C}_n^{\ast} {\rm e}^{-i \psi_n} \right)}, \label{SolAyW1}
\end{equation}
\begin{equation}
A_{z1} = - {\frac {i} {2}} \sum \limits_{n=1}^3 {\left( {\cal C}_n {\rm e}^{i \psi_n} - {\cal C}_n^{\ast} {\rm e}^{-i \psi_n} \right)}. \label{SolAzW1}
\end{equation}
The latter two equations imply that whistler waves are circularly polarized and this property expands on all other transverse field quantities and hydrodynamic variables in all orders. 

Having determined ${\cal A}_1$, the second of the first-order equations (\ref{PertMomBalvVn}) yields the following expression 
\begin{equation}
V_1 = \sum \limits_{n=1}^3 \Box_n {\cal C}_n {\rm e}^{i \psi_n}, \label{SolVW1}
\end{equation}
for the first-order transverse current velocity, where we have used the short-hand notation $\Box_n$, implying that $\Box_{\omega}$ should be taken for $\omega = \omega_n$. Similar notation will be used also for ${\cal L}_{\omega}$. 

It is worthwhile to emphasize that a remarkable feature of our description so far is the fact that whistler waves do not perturb the initial uniform density distribution $\varrho_0 = 1$ of plasma electrons. As we shall see, this property remains valid in second order as well. The plasma response to the induced whistler waves consists in a transverse velocity redistribution which follows exactly the behaviour of the whistlers.

\subsection{\label{subsec:second}Second Order}

The nonzero auxiliary perturbation quantities in second order are 
\begin{equation}
W_2 = {\frac {1} {2}} {\left( V_1 \partial_x {\cal A}_1^{\ast} + V_1^{\ast} \partial_x {\cal A}_1 \right)}, \label{AuxW2}
\end{equation}
\begin{equation}
U_2 = \partial_{t_1} {\cal A}_1 - \partial_{t_1} V_1, \quad \quad S_2 = - 2 {\left( \partial_x \partial_{x_1} - \partial_t \partial_{t_1} \right)} {\cal A}_1. \label{AuxUandS2}
\end{equation}
The solution of Eq. (\ref{BasicLongn}) in second order can be expressed as 
\begin{equation}
A_2 = - {\frac {k} {2}} \sum \limits_{m \neq n} {\frac {\Box_m + \Box_n} {{\left( \omega_m - \omega_n \right)} {\cal D}_0 {\left( \omega_m - \omega_n \right)}}} {\cal C}_m {\cal C}_n^{\ast} {\rm e}^{i {\left( \psi_m - \psi_n \right)}}, \label{SolAL2}
\end{equation}
where 
\begin{equation}
{\cal D}_0 {\left( \omega \right)} = \omega^2 - 1. \label{LongDisConst}
\end{equation}
Solving the first of Eqs. (\ref{PertMomBalvVn}) with the above expression for $A_2$ in hand, for the second-order longitudinal current velocity we obtain 
\begin{eqnarray}
v_2 = && - {\frac {k} {2}} \sum \limits_{m \neq n} {\frac {\Box_m + \Box_n} {\omega_m - \omega_n}} \nonumber 
\\ 
&& \times {\left[ 1 + {\frac {1} {{\cal D}_0 {\left( \omega_m - \omega_n \right)}}} \right]} {\cal C}_m {\cal C}_n^{\ast} {\rm e}^{i {\left( \psi_m - \psi_n \right)}}. \label{SolvL2}
\end{eqnarray}
Since $v_2$ does not depend on $x$, from the second-order Eq. (\ref{PertContn}) it follows that the second-order electron density vanishes 
\begin{equation}
\varrho_2 = 0. \label{SolRho2}
\end{equation}

A careful inspection of the second-order Eq. (\ref{BasicWhistlern}) shows that the source term on its right-hand-side would eventually produce secular terms. In order to avoid such terms, we require that the above-mentioned source term vanishes identically. The latter gives rise to the first order amplitude equations 
\begin{equation}
{\left( \partial_{t_1} + v_{gn} \partial_{x_1} \right)} {\cal C}_n = 0, \label{FirOrdAmpEq}
\end{equation}
where 
\begin{equation}
v_{gn} = {\frac {2 k {\cal L}_n} {2 \omega_n {\cal L}_n + \Box_n - 1}}, \label{GroupVel}
\end{equation}
is the $n$-th mode whistler wave group velocity \cite{TzenovTUBES}. 

Having eliminated the secular terms, it follows that the second-order transverse component of the electromagnetic vector potential vanishes 
\begin{equation}
{\cal A}_2 = 0. \label{SolAW2}
\end{equation}

\subsection{\label{subsec:third}Third Order - Derivation of the Amplitude Equations}

The nonzero contributions to the auxiliary perturbation quantities in third order can be expressed as 
\begin{equation}
R_3 = \partial_{x_1} v_2, \label{AuxR3}
\end{equation}
\begin{equation}
W_3 = \partial_{t_1} v_2 + {\frac {1} {2}} {\left( V_2 \partial_x {\cal A}_1^{\ast} + V_1 \partial_{x_1} {\cal A}_1^{\ast} + c.c. \right)}, \label{AuxW3}
\end{equation}
\begin{eqnarray}
U_3 = - {\frac {1} {2}} && \partial_t {\left( V_1 {\left| V_1 \right|}^2 \right)} - \partial_{t_1} V_2 - \partial_{t_2} V_1 \nonumber
\\ 
&& + \partial_{t_1} {\cal A}_2 + \partial_{t_2} {\cal A}_1 + v_2 \partial_x {\left( {\cal A}_1 - V_1 \right)}, \label{AuxU3}
\end{eqnarray}
\begin{equation}
N_3 = - 2 {\left( \partial_x \partial_{x_1} - \partial_t \partial_{t_1} \right)} A_2, \label{AuxN3}
\end{equation}
\begin{equation}
S_3 = - {\left( \partial_{x_1}^2 + 2 \partial_x \partial_{x_2} - \partial_{t_1}^2 - 2 \partial_t \partial_{t_2} \right)} {\cal A}_1, \label{AuxS3}
\end{equation}
where the notation ''$c.c.$'' stands for the complex conjugate counterpart. 

In the third-order Eq. (\ref{BasicWhistlern}), we retain only secular (resonant) terms, which follow the pattern of the three basic whistler wave modes (proportional to ${\rm e}^{i \psi_n}$). The rest contribute to the regular solution of the third order perturbation equations, involving higher harmonics and/or higher order combinations of the basic whistler modes. The condition for elimination of the above-mentioned secular contribution from the general perturbation solution of our initial system yields the sought for amplitude equations. Omitting straightforwardly reproducible calculation's details, we write down the final result 
\begin{eqnarray}
i \partial_t {\cal C}_n && + i v_{gn} \partial_x {\cal C}_n = - {\frac {1} {2}} {\frac {{\rm d} v_{gn}} {{\rm d} k}} \partial_x^2 {\cal C}_n \nonumber 
\\ 
&& + \sum \limits_m \Pi_{mn} {\cal C}_n {\left| {\cal C}_m \right|}^2 + \sum \limits_{m \neq n} \Gamma_{mn} {\cal C}_n {\left| {\cal C}_m \right|}^2, \label{BasicNSEWhistler}
\end{eqnarray}
where 
\begin{equation}
{\frac {{\rm d} v_{gn}} {{\rm d} k}} = {\frac {2} {1 - \Box_n - 2 \omega_n {\cal L}_n}} {\left[ v_{gn}^2 {\left( {\cal L}_n + {\frac {1 - \Box_n} {{\cal L}_n}} \right)} - {\cal L}_n \right]}, \label{DerivGrVel}
\end{equation}
\begin{equation}
\Pi_{mn} = {\frac {\omega_n \Box_m^2 \Box_n} {1 - \Box_n - 2 \omega_n {\cal L}_n}}, \label{CoeffPimn}
\end{equation}
\begin{eqnarray}
&& \Gamma_{mn} = {\frac {k^2} {2 {\left( 1 - \Box_n - 2 \omega_n {\cal L}_n \right)}}} \nonumber 
\\ 
&& \times {\frac {{\left( \Box_m + \Box_n \right)} {\left( 1 - \Box_m \right)}} {\omega_m - \omega_n}} {\left[ 1 + {\frac {1} {{\cal D}_0 {\left( \omega_m - \omega_n \right)}}} \right]}, \label{CoeffGammamn}
\end{eqnarray}
As already mentioned in the beginning of the present Section, the formal small parameter $\epsilon$ has been set equal to one, so that $x_1 = x$, $x_2 = x$ and $t_2 = t$.

Equations (\ref{BasicNSEWhistler}) comprise a system of three coupled nonlinear Schrodinger equations for the envelopes ${\cal C}_n$ of the three whistler wave modes. They describe the evolution of the slowly varying amplitudes of the generated transverse whistler wakefield. Since terms with $m=n$ are excluded from the second sum on the right-hand-side of Eq. (\ref{BasicNSEWhistler}), the matrix of coupling coefficients $\Gamma_{mn}$ represents a sort of a selection rule, according to which a generic mode $n$ cannot couple with itself. Note that this feature is a consequence of the vector character of the nonlinear coupling between modes and is due to the nonlinear Lorentz force. The first term (not present in the non relativistic case) involving the coupling matrix $\Pi_{mn}$ allows self-coupling and is entirely due to the relativistic character of the motion. Therefore, for a given mode $n$ the simplest nontrivial reduction of the coupled nonlinear Schrodinger equations consists of minimum two coupled equations in both the relativistic and the non relativistic case.

\section{\label{sec:solcnse}Solution of the Coupled Nonlinear Schrodinger Equations in the Non Relativistic Case} 

In order to examine the selective coupling between whistler modes, we first consider the non relativistic regime, where self coupling of particular modes is absent ${\left( \Pi_{mn} = 0 \right)}$. The results thus obtained in the non relativistic case will then be compared with the full relativistic dynamics of the whistler waves ${\left( \Pi_{mn} \neq 0 \right)}$. Straightforward evaluation of the dispersion coefficients $v_{gn}^{\prime} = {\rm d} v_{gn} / {\rm d} k$ shows that in a relatively wide range of plasma parameters one of them, say $v_{g2}^{\prime}$ (depending on the numbering of the roots of the whistler waves dispersion equation) is several orders of magnitude smaller than the other two. Thus, in a good approximation, Eqs. (\ref{BasicNSEWhistler}) can be written explicitly as 
\begin{eqnarray}
i \partial_t {\cal C}_1 && + i v_{g1} \partial_x {\cal C}_1 = - {\frac {v_{g1}^{\prime}} {2}} \partial_x^2 {\cal C}_1 \nonumber 
\\
&& + {\left( \Pi_{11} {\left| {\cal C}_1 \right|}^2 + \Sigma_{21} {\left| {\cal C}_2 \right|}^2 + \Sigma_{31} {\left| {\cal C}_3 \right|}^2 \right)} {\cal C}_1, \label{NLSchrod1}
\end{eqnarray}
\begin{eqnarray}
i \partial_t {\cal C}_2 && + i v_{g2} \partial_x {\cal C}_2 \nonumber 
\\ 
&& = {\left( \Sigma_{12} {\left| {\cal C}_1 \right|}^2 + \Pi_{22} {\left| {\cal C}_2 \right|}^2 + \Sigma_{32} {\left| {\cal C}_3 \right|}^2 \right)} {\cal C}_2, \label{NLSchrod2}
\end{eqnarray}
\begin{eqnarray}
i \partial_t {\cal C}_3 && + i v_{g3} \partial_x {\cal C}_3 = - {\frac {v_{g3}^{\prime}} {2}} \partial_x^2 {\cal C}_3 \nonumber 
\\
&& + {\left( \Sigma_{13} {\left| {\cal C}_1 \right|}^2 + \Sigma_{23} {\left| {\cal C}_2 \right|}^2 + \Pi_{33} {\left| {\cal C}_3 \right|}^2 \right)} {\cal C}_3, \label{NLSchrod3}
\end{eqnarray}
where 
\begin{equation}
\Sigma_{mn} = \Pi_{mn} + \Gamma_{mn}, \qquad \quad m \neq n. \label{CoupConst}
\end{equation}
Equation (\ref{NLSchrod2}) possesses a simple solution of the form 
\begin{equation}
{\cal C}_2 = g_2 {\rm e}^{-i \Psi {\left( x; t \right)}}, \label{SolNLSchrod2}
\end{equation}
where $g_2$ is a constant, while the phase $\Psi$ satisfies the equation 
\begin{equation}
\partial_t \Psi + v_{g2} \partial_x \Psi = \Sigma_{12} {\left| {\cal C}_1 \right|}^2 + \Pi_{22} g_2^2 + \Sigma_{32} {\left| {\cal C}_3 \right|}^2. \label{PhaseSchrod2}
\end{equation}
All of the above imply that our initial system (\ref{NLSchrod1}) -- (\ref{NLSchrod3}) can be reduced to a simpler system of two coupled nonlinear Schrodinger equations 
\begin{eqnarray}
i \partial_t {\cal C}_1 + && i v_{g1} \partial_x {\cal C}_1 = - {\frac {v_{g1}^{\prime}} {2}} \partial_x^2 {\cal C}_1 \nonumber 
\\ 
&& + {\left( \Pi_{11} {\left| {\cal C}_1 \right|}^2 + \Sigma_{21} g_2^2 + \Sigma_{31} {\left| {\cal C}_3 \right|}^2 \right)} {\cal C}_1, \label{Schrod1}
\end{eqnarray}
\begin{eqnarray}
i \partial_t {\cal C}_3 + && i v_{g3} \partial_x {\cal C}_3 = - {\frac {v_{g3}^{\prime}} {2}} \partial_x^2 {\cal C}_3 \nonumber 
\\ 
&& + {\left( \Sigma_{13} {\left| {\cal C}_1 \right|}^2 + \Sigma_{23} g_2^2 + \Pi_{33} {\left| {\cal C}_3 \right|}^2 \right)} {\cal C}_3. \label{Schrod3}
\end{eqnarray}

Following Ref. \citenum{TzenovPoP}, we first introduce new independent variables according to the relations  
\begin{equation}
\xi = - {\mathcal a} {\left( x - v_{g1} t \right)}, \quad \eta = {\mathcal a} {\left( x - v_{g3} t \right)}, \quad {\mathcal a} ={\frac {1} {v_{g1} - v_{g3}}}, \label{NewVariab}
\end{equation}
and rewrite Eqs. (\ref{Schrod1}) and (\ref{Schrod3}) as 
\begin{eqnarray}
i \partial_{\eta} {\cal C}_1 && = - {\frac {v_{g1}^{\prime} {\mathcal a}^2} {2}} {\left( \partial_{\xi} - \partial_{\eta} \right)}^2 {\cal C}_1 \nonumber 
\\
&& + {\left( \Pi_{11} {\left| {\cal C}_1 \right|}^2 + \Sigma_{21} g_2^2 + \Sigma_{31} {\left| {\cal C}_3 \right|}^2 \right)} {\cal C}_1, \label{NSchrod1}
\end{eqnarray}
\begin{eqnarray}
i \partial_{\xi} {\cal C}_3 && = - {\frac {v_{g3}^{\prime} {\mathcal a}^2} {2}} {\left( \partial_{\xi} - \partial_{\eta} \right)}^2 {\cal C}_3 \nonumber 
\\ 
&& + {\left( \Sigma_{13} {\left| {\cal C}_1 \right|}^2 + \Sigma_{23} g_2^2 + \Pi_{33} {\left| {\cal C}_3 \right|}^2 \right)} {\cal C}_3. \label{NSchrod3}
\end{eqnarray}
The next step is to seek traveling wave solutions through the standard ansatz
\begin{equation}
{\cal C}_1 = {\rm e}^{i {\left( \mu_1 \xi + \mu_2 \eta \right)}} {\cal P}_1 {\left( \eta  \right)}, \qquad {\cal C}_3 = {\rm e}^{i \mu_3 {\left( \xi + \eta \right)}} {\cal P}_3 {\left( \eta  \right)}. \label{StandAnsatz}
\end{equation}
Setting 
\begin{equation}
\mu_2 = \mu_1 - {\frac {1} {v_{g1}^{\prime} {\mathcal a}^2}}, \label{RelMu2Mu1}
\end{equation}
the latter allows us to transform the system of partial differential equations (\ref{NSchrod1}) and (\ref{NSchrod3}) into a system of second order ordinary differential equations as follows 
\begin{equation}
{\frac {{\rm d}^2 {\cal P}_1} {{\rm d} \eta^2}} + \nu_1^2 {\cal P}_1 = {\frac {2 \Pi_{11}} {v_{g1}^{\prime} {\mathcal a}^2}} {\cal P}_1^3 + {\frac {2 \Sigma_{31}} {v_{g1}^{\prime} {\mathcal a}^2}} {\cal P}_3^2 {\cal P}_1, \label{OrdSchrod1}
\end{equation}
\begin{equation}
{\frac {{\rm d}^2 {\cal P}_3} {{\rm d} \eta^2}} + \nu_3^2 {\cal P}_3 = {\frac {2 \Sigma_{13}} {v_{g3}^{\prime} {\mathcal a}^2}} {\cal P}_1^2 {\cal P}_3 + {\frac {2 \Pi_{33}} {v_{g3}^{\prime} {\mathcal a}^2}} {\cal P}_3^3, \label{OrdSchrod3}
\end{equation}
where 
\begin{equation}
\nu_1^2 = - {\frac {2} {v_{g1}^{\prime} {\mathcal a}^2}} {\left( \mu_1 - {\frac {1} {2 v_{g1}^{\prime} {\mathcal a}^2}} + g_2^2 \Sigma_{21} \right)}, \label{Freque1}
\end{equation}
\begin{equation}
\nu_3^2 = - {\frac {2} {v_{g3}^{\prime} {\mathcal a}^2}} {\left( \mu_3 + g_2^2 \Sigma_{23} \right)}. \label{Freque3}
\end{equation}
Note that having eliminated terms containing first order derivatives, we can now consider the wave amplitudes ${\cal P}_1$ and ${\cal P}_3$ real.

Since the dispersion coefficient $v_{g3}^{\prime}$ is generally negative for a wide range of plasma parameters, depending of the choice of the free parameters $\mu_1$ and $\mu_3$, one has the following two basic cases: 
\begin{itemize}
 \item the frequencies $\nu_1$ and $\nu_3$ are real provided 
\begin{equation}
\mu_3 > 0, \qquad \mu_1 < {\frac {1} {2 v_{g1}^{\prime} {\mathcal a}^2}} - g_2^2 \Sigma_{21}, \label{Case1}
\end{equation}
 \item the frequencies $\nu_1$ and $\nu_3$ are imaginary provided 
\begin{equation}
\mu_3 < 0, \qquad \mu_1 > {\frac {1} {2 v_{g1}^{\prime} {\mathcal a}^2}} - g_2^2 \Sigma_{21}, \label{Case2}
\end{equation}
\end{itemize} 
as well as cases involving obvious alternative combinations between $\mu_1$ and $\mu_3$. Here, we shall examine in more detail only the first case. 
\begin{figure}
\begin{center} 
\includegraphics[width=8.0cm]{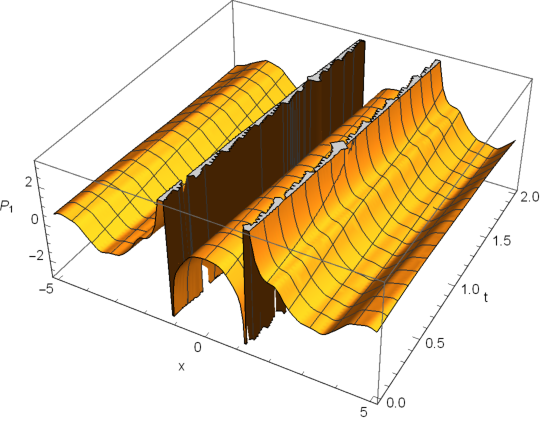}
\caption{\label{fig1:epsart} Evolution of the non relativistic traveling wave amplitude ${\cal P}_1$ for the case, where $\omega_e = 1$ $k = 1$, $\mu_1 = -1$, $\mu_3 = 1$, $g_1 = g_3 = 1$ and $g_2 = 0$.}
\end{center}
\end{figure}

The system of nonlinearly coupled Duffing equations (\ref{OrdSchrod1}) and (\ref{OrdSchrod3}) can be solved by utilizing a non-conventional method known as the method of formal series of Dubois-Violette \cite{TzenovBOOK,TzenovPoP}. The basic idea of this technique is to represent the solution of a generic nonlinear equation as a ratio of two formal Volterra series in powers of a (formal) perturbation parameter, rather than a conventional power series provided by standard perturbation theory. More generally, the method of formal series can be regarded as a nonlinear generalization of the Cramer's rule, wherein the solution of a linear system of equations is represented as a quotient of two determinants. Omitting details of the calculation procedure, which can be found in Ref. \citenum{TzenovPoP}, we state here only the final result 

\begin{figure}
\begin{center} 
\includegraphics[width=8.0cm]{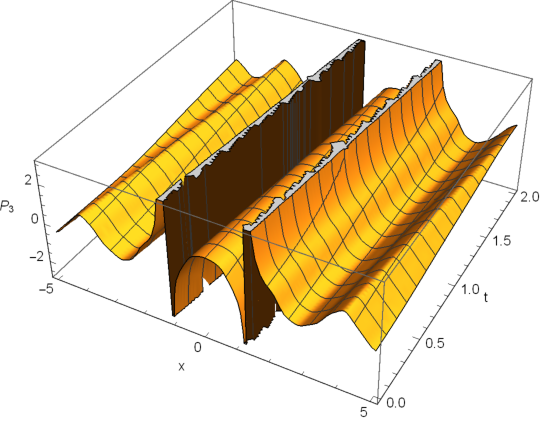}
\caption{\label{fig2:epsart} Evolution of the non relativistic traveling wave amplitude ${\cal P}_3$ for the case, where $\omega_e = 1$ $k = 1$, $\mu_1 = -1$, $\mu_3 = 1$, $g_1 = g_3 = 1$ and $g_2 = 0$.}
\end{center}
\end{figure}

\begin{equation}
{\cal P}_m {\left( \eta \right)} = {\frac {\sum \limits_{n=0}^{\infty} {\cal P}_m^{(n)} {\left( \eta \right)}} {\sum \limits_{n=0}^{\infty} D^{(n)} {\left( \eta \right)}}}, \qquad \quad m = 1, 3. \label{SolutDubViol}
\end{equation}
To second order, the terms in the above expression can be expressed as 
\begin{equation}
{\cal P}_1^{(0)} {\left( \eta \right)} = {\cal P}_{10} {\left( \eta \right)}, \quad {\cal P}_3^{(0)} {\left( \eta \right)} = {\cal P}_{30} {\left( \eta \right)}, \quad D^{(0)} = 1, \label{Terms1}
\end{equation}
\begin{eqnarray}
{\cal P}_1^{(1)} {\left( \eta \right)} = && \int \limits_0^{\eta} {\rm d} \sigma \sin \nu_1 {\left( \eta - \sigma \right)} \nonumber 
\\ 
&& \times {\left[ G_{11} {\cal P}_{10}^2 {\left( \sigma \right)} + G_{31} {\cal P}_{30}^2 {\left( \sigma \right)} \right]} {\cal P}_{10} {\left( \sigma \right)}, \label{Terms2}
\end{eqnarray}
\begin{eqnarray}
{\cal P}_3^{(1)} {\left( \eta \right)} && = \int \limits_0^{\eta} {\rm d} \sigma \sin \nu_3 {\left( \eta - \sigma \right)} \nonumber 
\\
&& \times {\left[ G_{13} {\cal P}_{10}^2 {\left( \sigma \right)} + G_{33} {\cal P}_{30}^2 {\left( \sigma \right)} \right]} {\cal P}_{30} {\left( \sigma \right)}, \label{Terms3}
\end{eqnarray}
\begin{equation}
D^{(1)} = 0, \qquad D^{(2)} = D_1^{(2)} + D_2^{(2)} + D_3^{(2)}, \label{Terms4}
\end{equation}
\begin{eqnarray}
D_1^{(2)} {\left( \eta \right)} && = - {\frac {1} {2}} \int \limits_0^{\eta} {\rm d} \lambda_1 \int \limits_0^{\eta} {\rm d} \lambda_2 \sin^2 \nu_1 {\left( \lambda_2 - \lambda_1 \right)} \nonumber 
\\
&& \times {\left[ 3 G_{11} {\cal P}_{10}^2 {\left( \lambda_1 \right)} + G_{31} {\cal P}_{30}^2 {\left( \lambda_1 \right)} \right]} \nonumber 
\\ 
&& \times {\left[ 3 G_{11} {\cal P}_{10}^2 {\left( \lambda_2 \right)} + G_{31} {\cal P}_{30}^2 {\left( \lambda_2 \right)} \right]}, \label{Terms5}
\end{eqnarray}
\begin{eqnarray}
&& D_2^{(2)} {\left( \eta \right)} = - 4 G_{13} G_{31} \int \limits_0^{\eta} {\rm d} \lambda_1 \int \limits_0^{\eta} {\rm d} \lambda_2 \sin \nu_1 {\left( \lambda_2 - \lambda_1 \right)} \nonumber 
\\
&& \times \sin \nu_3 {\left( \lambda_2 - \lambda_1 \right)} {\cal P}_{10} {\left( \lambda_1 \right)} {\cal P}_{30} {\left( \lambda_1 \right)} {\cal P}_{10} {\left( \lambda_2 \right)} {\cal P}_{30} {\left( \lambda_2 \right)}, \label{Terms6}
\end{eqnarray}
\begin{eqnarray}
D_3^{(2)} {\left( \eta \right)} && = - {\frac {1} {2}} \int \limits_0^{\eta} {\rm d} \lambda_1 \int \limits_0^{\eta} {\rm d} \lambda_2 \sin^2 \nu_3 {\left( \lambda_2 - \lambda_1 \right)} \nonumber 
\\
&& \times {\left[ G_{13} {\cal P}_{10}^2 {\left( \lambda_1 \right)} + 3 G_{33} {\cal P}_{30}^2 {\left( \lambda_1 \right)} \right]} \nonumber 
\\ 
&& \times {\left[ G_{13} {\cal P}_{10}^2 {\left( \lambda_2 \right)} + 3 G_{33} {\cal P}_{30}^2 {\left( \lambda_2 \right)} \right]}, \label{Terms7}
\end{eqnarray}
where 
\begin{equation}
{\cal P}_{10} = g_1 \cos {\left( \nu_1 \eta + {\mathcal h}_1 \right)}, \quad {\cal P}_{30} = g_3 \cos {\left( \nu_3 \eta + {\mathcal h}_3 \right)}. \label{Homogen}
\end{equation}
and 
\begin{equation}
G_{31} = {\frac {2 \Sigma_{31}} {v_{g1}^{\prime} {\mathcal a}^2 \nu_1}}, \qquad G_{13} = {\frac {2 \Sigma_{13}} {v_{g3}^{\prime} {\mathcal a}^2 \nu_3}}, \label{NewCoupCoeff1}
\end{equation}
\begin{equation}
G_{11} = {\frac {2 \Pi_{11}} {v_{g1}^{\prime} {\mathcal a}^2 \nu_1}}, \qquad G_{33} = {\frac {2 \Pi_{33}} {v_{g3}^{\prime} {\mathcal a}^2 \nu_3}}. \label{NewCoupCoeff2}
\end{equation}

The evolution of the whistler wave amplitudes ${\cal P}_1$ and ${\cal P}_3$ in the non relativistic case represented by Eq. (\ref{SolutDubViol}) up to second order is shown in Figures \ref{fig1:epsart} and \ref{fig2:epsart}. These imply that the second order traveling wave solution represents $1 / \eta$-damped quasi-periodic oscillations of the whistler mode amplitudes, which are practically unaltered in time. The solitary-like wave crests (positive as well as negative) with respect to the spatial variable for both ${\cal P}_1$ and ${\cal P}_3$ are almost monolithic structures over time and are symmetrical about the plane $x = 0$. 
\begin{figure}
\begin{center} 
\includegraphics[width=8.0cm]{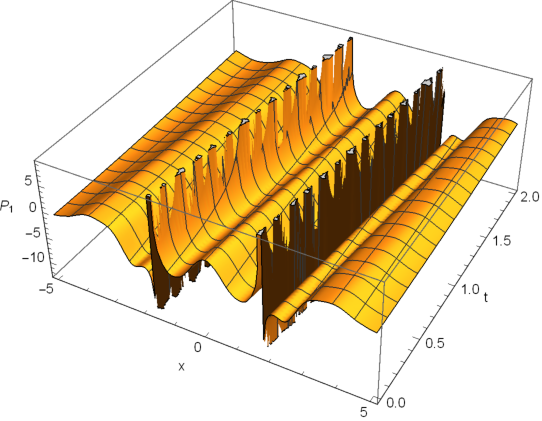}
\caption{\label{fig3:epsart} Evolution of the fully relativistic traveling wave amplitude ${\cal P}_1$ for the case, where $\omega_e = 1$ $k = 1$, $\mu_1 = -1$, $\mu_3 = 1$, $g_1 = g_3 = 1$ and $g_2 = 0$.}
\end{center}
\end{figure}

\begin{figure}
\begin{center} 
\includegraphics[width=8.0cm]{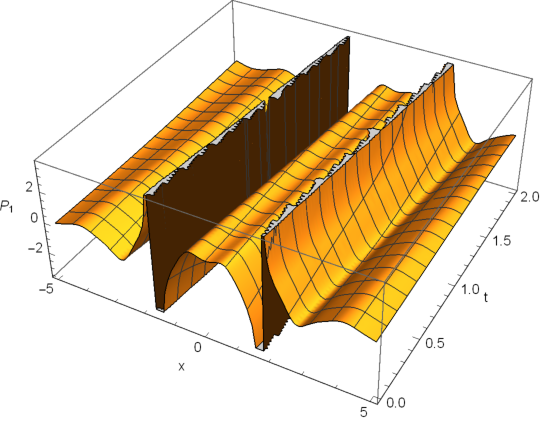}
\caption{\label{fig4:epsart} Evolution of the fully relativistic traveling wave amplitude ${\cal P}_3$ for the case, where $\omega_e = 1$ $k = 1$, $\mu_1 = -1$, $\mu_3 = 1$, $g_1 = g_3 = 1$ and $g_2 = 0$.}
\end{center}
\end{figure}

Figures \ref{fig3:epsart} and \ref{fig4:epsart} represent the fully relativistic case, for which the contribution of the mode self coupling terms $\Pi_{mm}$ has been taken into account. Note that the solitary-like wave crests with respect to the spatial variable for the mode amplitude ${\cal P}_1$ become time modulated, while those corresponding to ${\cal P}_3$ do not change their qualitative behaviour. 

Leading-order analysis \cite{Tabor} of the system of nonlinearly coupled Duffing equations (\ref{OrdSchrod1}) and (\ref{OrdSchrod3}) confirms that their asymptotic solution in the vicinity of a movable singularity scales as $1 / \eta$. As mentioned above, the formal series solution (\ref{SolutDubViol}) truncated at second order correctly reproduces the behaviour near this first-order pole. Note that the fading away peculiarities of this solution hold valid in all consecutive higher than the second one orders. In principle, one could proceed further and take into account higher order contributions, but qualitatively the demeanour of the solution will not change much. In addition, explicit calculations become rather cumbersome in fourth order and above, so that computerized analytical manipulations are mandatory to be employed.

According to Eq. (\ref{SolVW1}), the plasma response to the induced whistler waves consists in transverse velocity redistribution, which follows exactly the nonlinear behaviour of the whistlers. This means that the electron current flow is well confined ans localized in the transverse direction, such that on a scale $3 \div 4 \; c / \omega_e$ the tails of the electron density distribution can be considered as practically completely subdued. 

\section{\label{sec:conclude}Concluding Remarks} 

Starting from first principle and utilizing a technique known as the hydrodynamic substitution, a relativistic hydrodynamic system of equations describing the dynamics of various species in a cold quasi-neutral plasma immersed in and external solenoidal magnetic field has been obtained.

Based on the method of multiple scales, a further reduction of the macroscopic fluid and the wave equations for the self-consistent electromagnetic fields has been performed. This reduction represents by itself a separation of fast (oscillatory) variables from slow ones (called amplitudes or envelopes), which usually govern the formation of stable patterns on longer time and/or spatial scales. As a result of the analytic manipulations thus performed, a system comprising three coupled nonlinear Schrodinger equation for the three basic whistler modes has been derived. It is noteworthy to mention that our reduction has been performed in the single wave number $k$ mode approximation. It is however possible to carry out a full reduction, but the resultant amplitude equations comprise an infinite set of coupled nonlinear Schrodinger equations. In this case, one can think of a gas consisting of mutually interacting quasi-particles (whistlerons) \cite{Whistleron}. 

Using the method of formal series of Dubois-Violette, a traveling wave solution of the derived set of coupled nonlinear Schrodinger equations in the fully relativistic regime has been obtained. This solution is represented by a ratio of two formal Volterra series, and is not only compact and elegant but very useful for concrete practical applications. To provide a way of assessing higher order contributions, one needs as many terms in (\ref{SolutDubViol}) as possible. The calculations to obtain the fourth and higher order terms become rather cumbersome, so that computer-aided analytical manipulations are strictly necessary to be employed. 

An important feature of our description is that whistler waves do not modulate the initial uniform density of plasma electrons. The plasma response to the induced whistler waves consists in transverse velocity redistribution, which follows exactly the behaviour of the whistlers. The electron current flow is well localized in the transverse direction, such that on a spatial scale of $3 \div 4 \; c / \omega_e$ the tails of the electron density distribution can be considered as practically completely faded away. This property may have an important application for transverse focusing of charged particle beams in future laser plasma accelerators. Another interesting peculiarity are the selection rules in the non relativistic case governing the nonlinear mode coupling. According to these rules modes do not couple with themselves, which is a direct consequence of the vector character of the interaction.

We believe that the results obtained in the present article might have a wide class of possible applications ranging from laboratory experiments on laser plasma accelerators to observations of a variety of effects relevant to space plasmas.

\begin{acknowledgments}
The present work has been supported by Extreme Light Infrastructure -- Nuclear Physics (ELI-NP) Phase II, an innovative project co-financed by the Romanian Government and the European Union through the European Regional Development Fund. 
\end{acknowledgments}

\nocite{*}
\bibliography{aipsamp}

\end{document}